\documentclass[pra,onecolumn,floatfix,superscriptaddress,showpacs,amsfonts]{revtex4}
\usepackage{graphicx}
\usepackage{array}
\usepackage{amsthm}
\usepackage{amsmath}
\usepackage{amssymb}

\begin{document}
\title{Dynamics of cohering and decohering power under Markovian channels}

\author{Mingming Chen}
\affiliation{School of Mathematics and Information Science,Shaanxi Normal University, Xi¡¯an, 710062, China}

\author{Yu Luo}
\affiliation{College of Computer Science, Shaanxi Normal University, Xi'an, 710062, China}

\author{Lian-He Shao}
\affiliation{College of Computer Science, Shaanxi Normal University, Xi'an, 710062, China}

\author{Yongming Li}
\email{liyongm@snnu.edu.cn}
\affiliation{College of Computer Science, Shaanxi Normal University, Xi'an, 710062, China}

\begin{abstract}
In this paper, we investigate the cohering and decohering power for the one-qubit Markovian channels with respect to coherence in terms of the $l_{1}$-norm, the R$\acute{e}$nyi $\alpha$-relative entropy and the Tsallis $\alpha$-relative entropy. In the case of $\alpha=2$, the cohering and decohering power of the amplitude damping channel, the phase damping channel, the depolarizing channel, and the flip channels under the three measures of coherence are calculated analytically. The decohering power on the $x, y, z$ basis referring to the amplitude damping channel, the phase damping channel, the flip channel for every measure we investigated is equal. This property also happens in the cohering power of the phase damping channel, the depolarizing channel, and the flip channels. However, the decohering power of the depolarizing channel is independent to the reference basis, and the cohering power of the amplitude damping channel on the $x, y$ basis is different to that on the $z$ basis.
\end{abstract}
\maketitle

\section{Introduction}

Coherence, being the heart of interference phenomenon, arising from a critical property which is called quantum superposition. Coherence is one of the most important physical resource in quantum information processing, which can be used in quantum biology~\cite{Plenio08, Lloyd11, Levi14}, quantum thermodynamics~\cite{Rodr¡ä?guez-Rosario1308, Lostaglio15, Lostaglio1511, Brand?ao15, Narasimhachar15, C¡äwiklin¡äski15, Misra16}, and resource theories~\cite{Bromley15, Chitambar16, Du15, Girolami16, Napoli16, Rana16, Shao15, Winter16}. The importance of coherence encourages a lot of further studies on this subject~\cite{Liu16, Yao15, Du15, Cheng15, Bera15, Shao15, Streltsov15, Xi15, Zhang16, Chen1601, Situ}.

Recently, Baumgratz et.al. introduce a rigorous framework for quantifying coherence. A coherence measure satisfies four necessary criterias~\cite{Baumgratz14}. Considering a quantum state in a finite dimensional Hilbert space $\mathcal{H}$ with $d=dim(H)$, we note that $\mathcal{I}$ is a set of incoherent quantum states, which are diagonal in a set of fixed basis $\{|i\rangle\}^{d}_{i=1}$. Then any proper measure of the coherence $C$ must satisfy the following conditions:

(C1) $C(\rho)\geqslant0$ for all quantum states $\rho$ and $C(\rho)=0$ if and only if $\rho\in\mathcal{I}$;

(C2a) Monotonicity under all the incoherent completely positive and trace-preserving (ICPTP) maps $\Phi_{ICPTP}$: $C(\rho)\geqslant C(\Phi_{ICPTP}(\rho))$, where $\Phi_{ICPTP}(\rho)=K_n\rho K_n^\dagger$ and ${K_{n}}$ is a set of Kraus operators, which satisfies $\sum_{n}K_n^\dagger K_n\subset \mathcal{I}$ and $K_n \mathcal{I} K_n^\dagger\subset \mathcal{I}$;

(C2b) Monotonicity for average coherence under subselection measurements: $C(\rho)\geqslant \sum_n p_n C(\rho_n) $ where $\rho_n= \frac{K_{n}\rho\ K_{n}^{\dagger}}{p_{n}}$ and $p_{n}= Tr(K_{n}\rho\ K_{n}^{\dagger})$ for all $\sum_{n}K_n^\dagger K_n\subset \mathcal{I}$ and $K_n \mathcal{I} K_n^\dagger\subset \mathcal{I}$;

(C3) Non-increasing under mixing of quantum states(convexity): $\sum_n p_n C(\rho_n)\geqslant C(\sum_n p_n \rho_n )$ for any ensemble $\{p_n,\rho_n\}$.

Moreover, Rastegin et.al. propose the extension condition C2b~\cite{Rastegin16}. The extension condition C2b can be represented as

(C2b') $\sum\limits_{n}p_{n}q_{n}C(\rho)\leqslant C(\rho)$, where $p_{n}=Tr(K_{n}\rho K^{\dagger}_{n})$, $q_{n}=Tr(K_{n}\delta K^{\dagger}_{n})$, and $\delta$ is the nearest incoherent state to $\rho$.
\newline Considering the properties mentioned above, varies of coherence measures have been discussed, such as $l_1$-norm of off-diagonal elements of the quantum states~\cite{Baumgratz14}, the relative entropy of coherence~\cite{Baumgratz14}, the coherence of formation~\cite{Winter16}, the R$\acute{e}$nyi $\alpha$-relative entropy~\cite{Chitambar1602} and the Tsallis $\alpha$-relative entropy~\cite{Rastegin16}.

Quantum coherence can be destroyed by noise in the open system. However, sometimes it can be frozen~\cite{Bromley15} or increased in some special kinds of channels~\cite{Hu16}. To quantifying the power of a channel for creating or destroying the coherence of input quantum states, Mani $et.$ $al.$ give the definition of the cohering and the decohering power of quantum channels~\cite{Mani15}, and Bu $et.$ $al.$ define the coherence breaking indices for incoherent quantum channels~\cite{Bu16}.

In this paper, we mainly study the dynamics of the cohering and the decohering power of Markovian channels under the $l_{1}$-norm, the R$\acute{e}$nyi $\alpha$-relative entropy and the Tsallis $\alpha$-relative entropy. For the difficulty in studying the general value of $\alpha$, we only consider the case $\alpha=2$. We find that the decohering power of the amplitude channel, the phase damping channel, and the depolarizing channel under the $l_{1}$-norm, the R$\acute{e}$nyi $2$-relative entropy and the Tsallis $2$-relative  entropy are increased monotonically with $p$ from 0 to 1. However, it increases for $p\in[0, \frac{1}{2}]$ and decreases for $p\in[\frac{1}{2}, 1]$ with respect to the flip channels, reaching a maximal value of 1 at $p=\frac{1}{2}$. Moreover, we find the maximum of the cohering power of the amplitude damping channel, the phase damping channel, and the flip channels under the three measures are the same to 1. Finally, we find the cohering power vanishes for the depolarizing channel with respect to arbitrary basis.

This paper is organized as follows. In Sec.~\ref{sec:P}, we introduce the $l_{1}$-norm, the R$\acute{e}$nyi $\alpha$-relative entropy and the Tsallis $\alpha$-relative entropy of coherence. Moreover, we recall the definitions of cohering and decohering power. In Sec.~\ref{sec:Channels}, we firstly calculate the range of the $l_{1}$-norm, the R$\acute{e}$nyi $\alpha$-relative entropy and the Tsallis $\alpha$-relative entropy. Then we get the expression of that three measures under $K$ basis in the case of $\alpha=2$. In the end, we study the cohering and decohering power of Markovian channels under the $l_{1}$-norm, the R$\acute{e}$nyi $2$-relative entropy and the Tsallis $2$-relative entropy. We summarize our results in Sec.~\ref{sec:conclusion}.

\section{Preliminaries}\label{sec:P}

In order to quantify the cohering and the decohering power of general qubit Markovian channels, we introduce three measurements of coherence ,the $l_{1}$-norm, the R$\acute{e}$nyi $\alpha$-relative entropy and the Tsallis $\alpha$-relative entropy.

The $l_{1}$-norm of coherence $C_{l_1}$ is defined as~\cite{Baumgratz14}
\begin{equation}\label{Eq:1}
C_{l_1}(\rho)=\min\limits_{\delta \in \mathcal{I}}D(\rho,\delta)=\sum\limits_{i\neq j}|\rho_{ij}|,
\end{equation}
where $D(\rho,\delta)=\|\rho-\delta\|_{1}=\sum\limits_{i, j}|\rho_{ij}-\delta_{ij}|$ denotes the $l_{1}$ matrix norm, which means the measure induced by the $l_{1}$-norm is based on the minimal distance of $\rho$ to a set of incoherent states $\delta$. And $\rho_{ij}$ is the off-diagonal element of a quantum state $\rho$.  The $l_{1}$-norm of coherence satisfies the conditions of C1, C2a, C2b and C3 which is proved by Baumgratz et.al., so the $l_{1}$-norm of coherence is a valid coherence measure.

The R$\acute{e}$nyi $\alpha$-relative entropy of coherence $C_{R_{\alpha}}$ is defined as~\cite{Chitambar1601, Chitambar1602}
\begin{equation}\label{Eq:R}
C_{R_{\alpha}}(\rho)=\min\limits_{\delta \in \mathcal{I}}S_{\alpha}(\rho\|\delta)=\frac{\alpha}{\alpha-1}\log\sum\limits_{i}(\langle i|\rho^{\alpha}|i\rangle)^{\frac{1}{\alpha}},
\end{equation}
\newline where $\alpha\in[0, 2]$. Note that $S_{\alpha}(\rho\|\delta)=\frac{1}{\alpha-1}\log Tr(\rho^{\alpha}\delta^{1-\alpha})$ for all $0\leqslant\alpha$ is the R$\acute{e}$nyi $\alpha$-relative entropy, and it reduces to the von Neumann relative entropy when $\alpha\rightarrow1$,i.e., $\lim\limits_{\alpha\rightarrow1} S_{\alpha}(\rho\|\delta)=S(\rho\|\delta)=Tr[\rho(\ln\rho-\ln\delta)]$. At the same time the $C_{R_{\alpha}}$ will reduces to the relative entropy of coherence $C_{r}$
\begin{equation}\label{Eq:CR}
C_{r}(\rho)=\min\limits_{\delta \in \mathcal{I}}S(\rho\|\delta)=S(\rho_{diag})-S(\rho).
\end{equation}
In Ref.\cite{Shao1609}, Shao $et.$ $al.$ show the R$\acute{e}$nyi $\alpha$-relative entropy of coherence violate the condition C2b and the extension condition C2b' for $\alpha\in(0,1)$, so we conclude that the measure of coherence induced by the R$\acute{e}$nyi $\alpha$-relative entropy is not a good measure for quantifying coherence. Due to the R$\acute{e}$nyi $\alpha$-relative entropy of coherence fulfills the condition C1 and C2a, so it can act as a coherence monotone.

The Tsallis $\alpha$-relative entropy of coherence $C_{T_{\alpha}}$ is defined as~\cite{Rastegin16}
\begin{equation}\label{Eq:T}
C_{T_{\alpha}}(\rho)=\min\limits_{\delta \in \mathcal{I}}D_{\alpha}(\rho\|\delta)=\frac{1}{\alpha-1}\left[\left(\sum\limits_{i}(\langle i|\rho^{\alpha}|i\rangle)^{\frac{1}{\alpha}}\right)^{\alpha}-1\right],
\end{equation}
\newline where $D_{\alpha}(\rho\|\delta)=\frac{Tr(\rho^{\alpha}\delta^{1-\alpha})-1}{\alpha-1}$ for $0<\alpha$ and $\alpha\neq1$ denotes the Tsallis relative $\alpha$ entropy, and it reduces to the von Neumann relative entropy when $\alpha\rightarrow1$,i.e., $\lim\limits_{\alpha\rightarrow1} D_{\alpha}(\rho\|\delta)=S(\rho\|\delta)=Tr[\rho(\ln\rho-\ln\delta)]$. At the same time, $C_{T_{\alpha}}$ will reduces to Eq.~\eqref{Eq:CR}. In Ref.\cite{Rastegin16}, the author proves that the Tsallis $\alpha$-relative entropy of coherence satisfies the conditions of C1, C2a and C3 for all $\alpha\in[0,2]$, but it may violate C2b in some situations. However, it satisfies an extension condition C2b'. We then conclude that the Tsallis $\alpha$-relative entropy of coherence can be used as a coherence measure.

Mani $et.$ $al.$ give the definition of the power of a quantum channel $\varepsilon$ for creating or destroying the coherence of input quantum states. The cohering power of a channel is the maximal amount of coherence that it creates when acting on a completely incoherent state. For any quantum channels $\varepsilon$, the cohering power is defined as~\cite{Mani15}
\begin{equation}\label{Eq:CP1}
C^K_{C}(\varepsilon)= \max\limits_{\rho\in\mathcal{I}}\{C^K(\varepsilon(\rho))-C^K(\rho)\}=\max\limits_{\rho\in\mathcal{I}}C^K(\varepsilon(\rho)).
\end{equation}
$C^K_{C}$ denotes the cohering power of the coherence measure $C$, $K$ is the reference basis, and we have used property (C1) in the second equation in Eq.\eqref{Eq:CP1}.

Using the convexity property (C3) of coherence measures, Eq.~\eqref{Eq:CP1} can be written in a simpler modality on the orthonormal basis vectors
\begin{equation}\label{Eq:CP2}
C^K_{C}(\varepsilon)=\max\limits_{i}C^K(\varepsilon(|k_{i}\rangle\langle k_{i}|)).
\end{equation}

The decohering power of the channel $\varepsilon$ is the maximum amount by which it reduces the coherence of a maximally coherent state. The decohering power for a quantum channel $\varepsilon$ is define as [2]
\begin{equation}\label{Eq:DP1}
D^K_{C}(\varepsilon)=\max\limits_{\rho\in M}\{C^K(\rho)-C^K(\varepsilon(\rho))\}.
\end{equation}
$D^K_{ C}$ denotes the decohering power of the coherence measure $C$, and $K$ is the reference basis. $M$ is a set of maximally coherent states. Because all maximally coherent states are pure ones, Eq.~\eqref{Eq:DP1} can be simplified as
\begin{equation}\label{Eq:DP2}
D^K_{C}(\varepsilon)=1-\min\limits_{\rho\in M} C^K(\varepsilon(\rho)).
\end{equation}
Note that the definitions of the cohering and the decohering power of quantum channels are valid for any types of coherence measures.

\section{Cohering and decohering power of Markovian channels}\label{sec:Channels}

As we all know that the coherence of input states may be changed by the quantum channels, then it is necessary
to measure the corresponding changes of the quantum coherence. For this purpose, we will study the power of the Markovian noisy one-qubit channels for creating or destorying the quantum coherence. Firstly, we introduce the notion of $K$ coherence, where $\mathbf{k}=(k_{x},k_{y},k_{z})$ is a unit vector standing for the reference $K$ basis $\{\frac{\mathbb{I}+\mathbf{k}\cdot\sigma}{2}, \frac{\mathbb{I}-\mathbf{k}\cdot\sigma}{2}\}$. For a general qubit
\begin{equation}\label{Eq:qubit}
\rho=\frac{1}{2}(\mathbb{I}+\mathbf{r}\cdot\sigma),
\end{equation}
where $\mathbf{r}=(r_{x},r_{y},r_{z})$ is a real vector which satisfies $\|\mathbf{r}\|\leqslant1$, and $\sigma=(\sigma_{x},\sigma_{y},\sigma_{z})$ is the vector of Pauli matrices.

Mani $et.$ $al.$ give the $l_{1}$-norm of coherence with respect to $K$ basis~\cite{Mani15}
\begin{equation}\label{Eq:1K}
C^K_{1}(\rho)=r\sqrt{1-(\mathbf{\hat{r}}\cdot\mathbf{k})^{2}},
\end{equation}
where $\hat{r}=\frac{\mathbf{r}}{r}$. In general, for any one-qubit state Eq.\eqref{Eq:1K} satisfies
\begin{equation}\label{Eq:FW1}
0\leqslant C^K_{1}(\rho)\leqslant1.
\end{equation}
In Ref.\cite{Shao1609}, Shao $et.$ $al.$ give the range of Eq.~\eqref{Eq:R} with respect to a set of basis $\{|i\rangle\}^{d}_{i=1}$. After simple calculation, we find Eq. ~\eqref{Eq:R} satisfies
\begin{equation}\label{Eq:FW2}
0\leqslant C^K_{R_{\alpha}}(\rho)\leqslant1+\log\left[2(1+\|\mathbf{r}\|^{2})\right],
\end{equation}
for all $\alpha\in[0, 2]$ under $K$ basis. Analogously, we derive the range of Eq.~\eqref{Eq:T} for a general qubit under $K$ basis referring to Ref.\cite{Rastegin16}
\begin{equation}\label{Eq:FW3}
0\leqslant C^K_{T_{\alpha}}(\rho)\leqslant-\ln_{\alpha}\frac{1}{4(1+\|\mathbf{r}\|^{2})},
\end{equation}
for $\alpha\in[0, 2]$, and
\begin{equation}\label{Eq:FW4}
0\leqslant C^K_{T_{\alpha}}(\rho)\leqslant\frac{1}{\alpha-1}\left[4\left(1+\|\mathbf{r}\|^{2})(1+\sqrt{4(1+\|\mathbf{r}\|^{2})-1}\right)^{\alpha-2}-1\right],
\end{equation}
for $\alpha\in[2, \infty]$, where $\ln_{\alpha}(x)=\frac{x^{1-\alpha}-1}{1-\alpha}$ is the $\alpha$ logarithm. The maximum value of Eq.~\eqref{Eq:FW1}, Eq.~\eqref{Eq:FW2}, Eq.~\eqref{Eq:FW3} and Eq.~\eqref{Eq:FW4} being achieved for a set of maximally coherence pure states with respect to $K$ basis,i.e., $|\psi\rangle=\frac{1}{\sqrt{2}}|\mathbf{k}_{+}\rangle+e^{i\Omega}|\mathbf{k}_{-}\rangle$, where $|\mathbf{k}_{\pm}\rangle$ is eigenvector of $\mathbf{k}\cdot\sigma$.

For the difficulty in calculating the expressions of the R$\acute{e}$nyi $\alpha$-relative entropy and the Tsallis $\alpha$-relative entropy for general $\alpha$ under $K$ basis, we execute a case study of $\alpha=2$. Substituting Eq.~\eqref{Eq:qubit} into Eq.~\eqref{Eq:R} and Eq.~\eqref{Eq:T} respectively, we obtain the R$\acute{e}$nyi $2$-relative entropy and the Tsallis $2$-relative entropy under $K$ basis as follows
\begin{equation}\label{Eq:R2K}
C^K_{R_{2}}(\rho)=2\log \left[\frac{1}{2}\left(\sqrt{1+\|\mathbf{r}\|^2+2\mathbf{r}\cdot\mathbf{k}}+\sqrt{1+\|\mathbf{r}\|^2-2\mathbf{r}\cdot\mathbf{k}}\right)\right],
\end{equation}
\begin{equation}\label{Eq:T2K}
C^K_{T_{2}}(\rho)=\left[\frac{1}{2}\left(\sqrt{1+\|\mathbf{r}\|^2+2\mathbf{r}\cdot\mathbf{k}}+\sqrt{1+\|\mathbf{r}\|^2-2\mathbf{r}\cdot\mathbf{k}}\right)\right]^2-1.
\end{equation}

Next, we will study the cohering and decohering power of Markovian noisy one-qubit channels under the $l_{1}$-norm, the R$\acute{e}$nyi $2$-relative entropy and the Tsallis $2$-relative entropy in four parts.

\subsection{Amplitude damping channel}

To study the cohering and the decohering power of a channel under the $l_{1}$-norm, the R$\acute{e}$nyi $2$-relative entropy and the Tsallis $2$-relative entropy, let's start with the amplitude damping channel $\varepsilon_{ad}$, which is characterised by Kraus operators~\cite{Nielsen00}
\begin{equation}\label{Eq:AD}
K_1=\left(
  \begin{array}{lcr}
 1 & 0 \\
 0 & \sqrt{1-p}
  \end{array}
    \right),
K_2=\left(
  \begin{array}{lcr}
 0 & \sqrt{p} \\
 0 & 0
  \end{array}
    \right),
\end{equation}
where $p\in\left[0,1\right]$ is a parametrization of time $t$, with $p=0$ corresponding to $t=0$ and $p=1$ corresponding to $t\rightarrow\infty$. It transforms the input Bloch vector $\mathbf{r}=(r_{x},r_{y},r_{z})$ into
\begin{equation}\label{Eq:ADr}
\varepsilon_{ad}(\mathbf{r})=(\sqrt{1-p}r_{x},\sqrt{1-p}r_{y},p+r_{z}(1-p)).
\end{equation}

\begin{figure}\label{fig1}
  \includegraphics[scale=0.6]{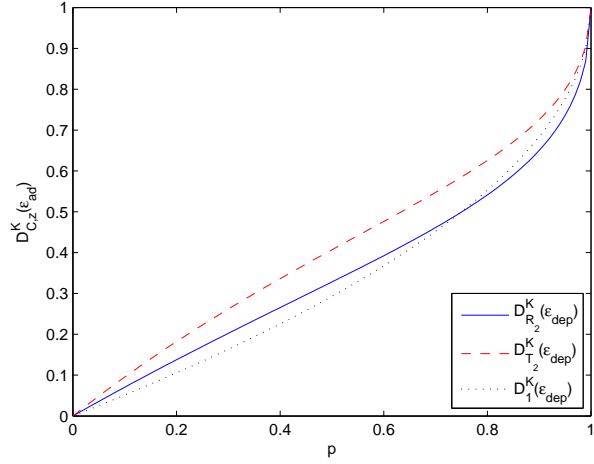}
\caption{(Color online) The decohering power of the amplitude damping channel with respect to $x, y, z$ basis. The black points line denotes the decohering power of the amplitude damping channel with respect to the $l_{1}$-norm on the $x, y, z$ basis. The red dotted line denotes the R$\acute{e}$nyi $2$-relative entropy, and the blue valid line denotes the Tsallis $2$-relative entropy.}.
\label{Fig_1}
\end{figure}

For the decohering power of the amplitude damping channel $\varepsilon_{ad}$, we take $\mathbf{r}\cdot\mathbf{k}=0$. Then taking Eq.~\eqref{Eq:ADr} into Eq.~\eqref{Eq:1K}, Eq.~\eqref{Eq:R2K} and Eq.~\eqref{Eq:T2K} respectively, we have the coherence of the maximally coherent states in terms of the $l_{1}$-norm, the Renyi $2$-relative entropy and the Tsallis relative $2$-entropy
\begin{equation}\label{Eq:Cad1}
C^K_{1}(\varepsilon_{ad}(\rho)=\sqrt{\mu-\nu^{2}},
\end{equation}
\begin{equation}\label{Eq:CadR2}
C^K_{R_{2}}(\varepsilon_{ad}(\rho)=2\log\left[\frac{1}{2}\left(\sqrt{1+\mu+2\nu}+\sqrt{1+\mu-2\nu}\right)\right],
\end{equation}
\begin{equation}\label{Eq:CadT2}
C^K_{T_{2},z}(\varepsilon_{ad}(\rho))=\left[\frac{1}{2}\left(\sqrt{1+\mu+2\nu}+\sqrt{1+\mu-2\nu}\right)\right]^2-1,
\end{equation}
\newline where $\mu=(p^{2}-p)r^{2}_{z}+2p(1-p)r_{z}+(p^{2}-p+1)$, and $\nu=k_{z}(r_{z}\sqrt{1-p}(\sqrt{1-p}-1)+p)$. Note that Eq.~\eqref{Eq:Cad1}, Eq.~\eqref{Eq:CadR2}, and Eq.~\eqref{Eq:CadT2} are the same for all $\rho\in M$. According to Eq.~\eqref{Eq:DP2}, we have
\begin{equation}\label{Eq:Dad1}
D^K_{1}(\varepsilon_{ad})=1-\min\limits_{\rho\in M}\sqrt{\mu-\nu^{2}},
\end{equation}
\begin{equation}\label{Eq:DadR2}
D^K_{R_{2}}(\varepsilon_{ad})=1-\min\limits_{\rho\in M}\left\{2\log\left[\frac{1}{2}\left(\sqrt{1+\mu+2\nu}+\sqrt{1+\mu-2\nu}\right)\right]\right\},
\end{equation}
\begin{equation}\label{Eq:DadT2}
D^K_{T_{2}}(\varepsilon_{ad})=2-\min\limits_{\rho\in M}\left[\frac{1}{2}\left(\sqrt{1+\mu+2\nu}+\sqrt{1+\mu-2\nu}\right)\right]^2.
\end{equation}
For the $x, y, z$ basis, $k_{z}$ is either 0 or 1, then the decohering power of the amplitude damping channel with respect to $x, y, z$ basis are
\begin{equation}\label{Eq:Dad1z}
D^K_{1,x}(\varepsilon_{ad})=D^K_{1,y}(\varepsilon_{ad})=D^K_{1,z}(\varepsilon_{ad})=1-\sqrt{1-p},
\end{equation}
\begin{equation}\label{Eq:DadR2z}
D^K_{R_{2},x}(\varepsilon_{ad})=D^K_{R_{2},y}(\varepsilon_{ad})=D^K_{R_{2},z}(\varepsilon_{ad})=1-2\log\left[\frac{1}{2}\left(\sqrt{p^2+p+2}+\sqrt{p^2-3p+2}\right)\right],
\end{equation}
\begin{equation}\label{Eq:DadT2z}
D^K_{T_{2},x}(\varepsilon_{ad})=D^K_{T_{2},y}(\varepsilon_{ad})=D^K_{T_{2},z}(\varepsilon_{ad})=2-\left[\frac{1}{2}\left(\sqrt{p^2+p+2}+\sqrt{p^2-3p+2}\right)\right]^2.
\end{equation}
As shown in Fig.~\ref{Fig_1}, we find that the decohering power with respect to the $x, y, z$ basis are increased monotonically with $p$ from 0 to maximal value 1.

In the next, we will calculate the cohering power. In this case, we take $\mathbf{r}=\pm\mathbf{k}$. Taking Eq.~\eqref{Eq:ADr} into Eq.~\eqref{Eq:1K}, Eq.~\eqref{Eq:R2K} and Eq.~\eqref{Eq:T2K} respectively, we have
\begin{equation}\label{Eq:Cad11}
C^K_{1}(\varepsilon_{ad}(\rho))=\sqrt{\mu-\omega^{2}},
\end{equation}
\begin{equation}\label{Eq:CadR21}
C^K_{R_{2}}(\varepsilon_{ad}(\rho))=2\log\left[\frac{1}{2}\left(\sqrt{1+\mu+2\omega}+\sqrt{1+\mu-2\omega}\right)\right],
\end{equation}
\begin{equation}\label{Eq:CadT21}
C^K_{T_{2},z}(\varepsilon_{ad}(\rho))=\left[\frac{1}{2}\left(\sqrt{1+\mu+2\omega}+\sqrt{1+\mu-2\omega}\right)\right]^2-1,
\end{equation}
where $\omega=k^2_{z}\sqrt{1-p}(\sqrt{1-p}-1)+k_{z}p+\sqrt{1-p}$. According to Eq.~\eqref{Eq:CP2}, we derive the cohering power of the amplitude damping channel under the $l_{1}$-norm, the R$\acute{e}$nyi $2$-relative entropy and the Tsallis $2$-relative entropy
\begin{equation}\label{Eq:Cad12}
C^K_{1}(\varepsilon_{ad})=\sqrt{\mu-\omega^{2}},
\end{equation}
\begin{equation}\label{Eq:CadR22}
C^K_{R_{2}}(\varepsilon_{ad})=2\log\left[\frac{1}{2}\left(\sqrt{1+\mu+2\omega}+\sqrt{1+\mu-2\omega}\right)\right],
\end{equation}
\begin{equation}\label{Eq:CadT22}
C^K_{T_{2},z}(\varepsilon_{ad})=\left[\frac{1}{2}\left(\sqrt{1+\mu+2\omega}+\sqrt{1+\mu-2\omega}\right)\right]^2-1.
\end{equation}
The maximal cohering power of the amplitude damping channel with respect to arbitrary basis under every measures studied is 1. For the $x, y$ basis, $k_{z}$ is 0, we have
\begin{equation}\label{Eq:Cad13}
C^K_{1,x}(\varepsilon_{ad})=C^K_{1,y}(\varepsilon_{ad})=p,
\end{equation}
\begin{equation}\label{Eq:CadR23}
C^K_{R_{2},x}(\varepsilon_{ad})=C^K_{R_{2},y}(\varepsilon_{ad})=2\log\left[\frac{1}{2}\left(\sqrt{p^2+(\sqrt{1-p}+1)^{2}}+\sqrt{p^2+(\sqrt{1-p}-1)^{2}}\right)\right],
\end{equation}
\begin{equation}\label{Eq:CadT23}
C^K_{T_{2},x}(\varepsilon_{ad})=C^K_{T_{2},y}(\varepsilon_{ad})=\left[\frac{1}{2}\left(\sqrt{p^2+(\sqrt{1-p}+1)^{2}}+\sqrt{p^2+(\sqrt{1-p}-1)^{2}}\right)\right]^2-1.
\end{equation}
The cohering power of the amplitude damping channel on the $x, y$ basis increased monotonically from 0 to 1. For the $z$ basis, $k_{z}$ is 1. Using Eq.~\eqref{Eq:CP2} again we have
\begin{equation}\label{Cad}
C^K_{1,z}(\varepsilon_{ad})=C^K_{R_{2},z}(\varepsilon_{ad})=C^K_{T_{2},z}(\varepsilon_{ad})=0.
\end{equation}
It means that the amplitude damping channel doesn't have any cohering power with respect ro $z$ basis,i.e, the amplitude damping channel $\varepsilon_{ad}$ is an incoherent channel on the $z$ basis.

\subsection{Phase damping channel}

A quantum channel with Kraus operators
\begin{equation}\label{Eq:PD}
K_1=\left(
  \begin{array}{lcr}
 1 & 0 \\
 0 & \sqrt{1-p}
  \end{array}
    \right),
K_2=\left(
  \begin{array}{lcr}
 0 & 0 \\
 0 & \sqrt{p}
  \end{array}
    \right),
\end{equation}
is called phase damping channel~\cite{Nielsen00}, denoted by $\varepsilon_{pd}$. It converts the input Bloch vector $\mathbf{r}=(r_{x},r_{y},r_{z})$ into
\begin{equation}\label{Eq:PDr}
\varepsilon_{pd}(\mathbf{r})=(\sqrt{1-p} r_{x},\sqrt{1-p} r_{y},r_{z}).
\end{equation}

\begin{figure}\label{fig_2}
  \includegraphics[scale=0.8]{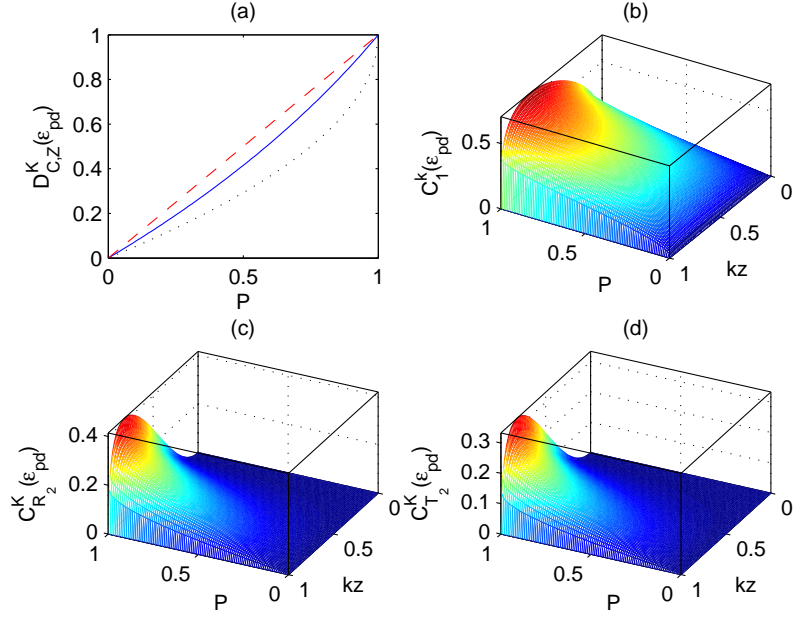}
\caption{(Color online) (a) The decohering power of phase damping channel with $p$ under the $l_{1}$-norm(black points line), the R$\acute{e}$nyi $2$-relative entropy(blue valid line), and the Tsallis $2$-relative entropy(red dotted line). (b)The cohering power of the phase damping channel under the $l_{1}$-norm, whose maximum is approximately equal to 0.7071. (c)The cohering power the phase damping channel under the R$\acute{e}$nyi $2$-relative entropy, whose maximum is approximately equal to 0.4150. (d)The cohering power of the phase damping channel under the Tsallis $2$-relative entropy, whose maximum is approximately equal to 0.3333.}.
\label{Fig_2}
\end{figure}

Substituting Eq.~\eqref{Eq:PDr} into Eq.~\eqref{Eq:1K}, Eq.~\eqref{Eq:R2K}, and Eq.~\eqref{Eq:T2K} respectively, and using Eq.~\eqref{Eq:DP2} we derive the decohering power of the phase damping channel with respect to the $l_{1}$-norm, the R$\acute{e}$nyi $2$-relative entropy and the Tsallis $2$-relative entropy
\begin{equation}\label{Eq:DPD1}
D^{K}_{1}(\varepsilon_{pd})=1-\min\limits_{\rho\in M}
\sqrt{p(r^{2}_{z}-1)-(1-\sqrt{1-p})(r_{z}k_{z})^{2}+1},
\end{equation}
\begin{equation}\label{Eq:DPDR}
D^{K}_{R_{2}}(\varepsilon_{pd})=1-\min\limits_{\rho\in M}\left\{2\log\left[\frac{1}{2}\left(\sqrt{2-p+pr^{2}_{z}+2(1-\sqrt{1-p})r_{z}k_{z}}+\sqrt{2-p+pr^{2}_{z}-2(1-\sqrt{1-p})r_{z}k_{z}}\right)\right]\right\},
\end{equation}
\begin{equation}\label{Eq:DPDT}
D^{K}_{T_{2}}(\varepsilon_{pd})=2-\min\limits_{\rho\in M}
\left[\frac{1}{2}\left(\sqrt{2-p+pr^{2}_{z}+2(1-\sqrt{1-p})r_{z}k_{z}}+\sqrt{2-p+pr^{2}_{z}-2(1-\sqrt{1-p})r_{z}k_{z}}\right)\right]^2.
\end{equation}
After some algebraic calculations, the decohering power of the phase damping channel with respect to the $x$, $y$, $z$ basis are
\begin{equation}\label{Eq:DPDxyz1}
D^{K}_{1,x}(\varepsilon_{pd})=D^{K}_{1,y}(\varepsilon_{pd})=D^{K}_{1,z}(\varepsilon_{pd})=1-\sqrt{1-p},
\end{equation}
\begin{equation}\label{Eq:DPDRxyzR}
D^{K}_{R_{2},x}(\varepsilon_{pd})=D^{K}_{R_{2},y}(\varepsilon_{pd})=D^{K}_{R_{2},z}(\varepsilon_{pd})=1-\log(2-p),
\end{equation}
\begin{equation}\label{Eq:DPDRxyzT}
D^{K}_{T_{2},x}(\varepsilon_{pd})=D^{K}_{T_{2},y}(\varepsilon_{pd})=D^{K}_{T_{2},z}(\varepsilon_{pd})=p.
\end{equation}
The change of the decohering power of the phase damping channel with respect to the $l_{1}$-norm, the R$\acute{e}$nyi $2$-relative entropy and the Tsallis $2$-relative entropy on the $x, y, z$ basis is in Fig.\ref{Fig_2}(a), from which we can see they all increased monotonically with $p$ from 0 to the maximal value 1.

For the cohering power of the phase damping channel, we take $\mathbf{r}=\pm\mathbf{k}$ as the reference basis which are converted by $\varepsilon_{pd}$
\begin{equation}\label{Eq:PDk}
\pm(k_{x},k_{y},k_{z})=\pm(\sqrt{1-p}k_{x}, \sqrt{1-p}k_{y}, k_{z}).
\end{equation}
Taking Eq.~\eqref{Eq:PDk} into Eq.~\eqref{Eq:1K}, Eq.~\eqref{Eq:R2K}, Eq.~\eqref{Eq:T2K} respectively, then using Eq.~\eqref{Eq:CP2} give the cohering power of the phase damping channel with respect to the $l_{1}$-norm, the $R\acute{e}nyi$ $2$-relative entropy and the Tsallis $2$-relative entropy
\begin{equation}\label{Eq:CPD12}
C^{K}_{1}(\varepsilon_{pd})=\sqrt{p(k^2_{z}-1)-((\sqrt{1-p}-1)k^2_{z}+\sqrt{1-p})^{2}+1},
\end{equation}
\begin{equation}\label{Eq:CPDR2}
C^{K}_{R_{2}}(\varepsilon_{pd})=2\log\left[\frac{1}{2}(\sqrt{\xi}\\+\sqrt{\eta}\right],
\end{equation}
\begin{equation}\label{Eq:CPDT2}
C^{K}_{T_{2}}(\varepsilon_{pd})=\left[\frac{1}{2}(\sqrt{\xi}\\+\sqrt{\eta}\right]^{2}-1,
\end{equation}
where $\xi=k_{z}^{2}(p+2(1-\sqrt{1-p}))+2(1+\sqrt{1-p})-p$, and $\eta=k^{2}_{z}(p-2(1-\sqrt{1-p}))+2(1-\sqrt{1-p})-p$. As shown in Fig.\ref{Fig_2}(b), (c) and (d), we have plotted the cohering power of the phase damping channel with respect to arbitrary basis under the three measures. The maximal cohering power of the phase damping channel with respect to the $l_{1}$-norm is 0.7071. While the maximal cohering power of the R$\acute{e}$nyi $2$-relative entropy and the Tsallis $2$-relative entropy are 0.4150 and 0.3333, respectively. For the $x$, $y$, $z$ basis, $k_{z}$ is either 0 or 1, so we have
\begin{equation}\label{Eq:CPDxyz1}
C^{K}_{1,x}(\varepsilon_{pd})=C^{K}_{1,y}(\varepsilon_{pd})=C^{K}_{1,z}(\varepsilon_{pd})=0,
\end{equation}
\begin{equation}\label{Eq:CPDxyzR}
C^{K}_{R_{2},x}(\varepsilon_{pd})=C^{K}_{R_{2},y}(\varepsilon_{pd})=C^{K}_{R_{2},z}(\varepsilon_{pd})=0,
\end{equation}
\begin{equation}\label{Eq:CPDxyzT}
C^{K}_{T_{2},x}(\varepsilon_{pd})=C^{K}_{T_{2},y}(\varepsilon_{pd})=C^{K}_{T_{2},z}(\varepsilon_{pd})=0.
\end{equation}
That is to say the phase damping channel has no cohering power on the $x, y, z$ basis.

\subsection{Depolarizing channel}

In this section, we will consider a dynamical evolution of a general state $\rho$ under the depolarizing channel $\varepsilon_{dep}$, which is acting as~\cite{Nielsen00}
\begin{equation}\label{Eq:DEPr}
\varepsilon_{dep}(\rho)=(1-p)\rho+p\frac{I}{2}.
\end{equation}

\begin{figure}\label{fig_3}
  \includegraphics[scale=0.6]{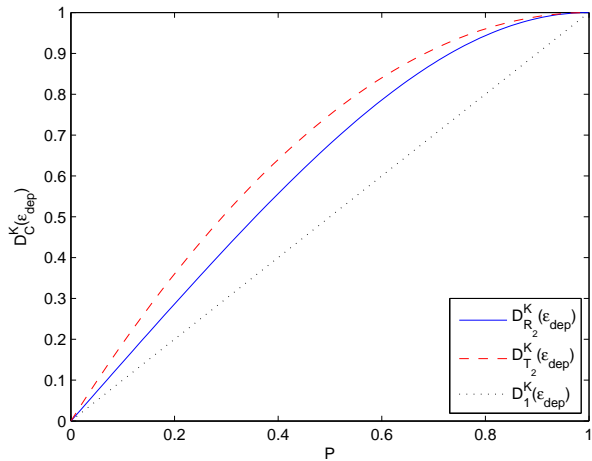}
\caption{(Color online) The decohering power of the depolarizing channel with respect to the $x, y, z$ basis. The black points line denotes the decohering power of the depolarizing channel with respect to the $l_{1}$-norm. The red dotted line denotes the R$\acute{e}$nyi $2$-relative entropy, and the blue valid line denotes Tsallis $2$-relative entropy.}.
\label{Fig_3}
\end{figure}

Taking advantage of Eq.~\eqref{Eq:1K}, Eq.~\eqref{Eq:R2K}, Eq.~\eqref{Eq:T2K}, and Eq.~\eqref{Eq:DP2}, we have the decohering power of the depolarizing channel
\begin{equation}\label{Eq:DDEP1}
D^{K}_{1}(\varepsilon_{dep})=p,
\end{equation}
\begin{equation}\label{Eq:DDEPR}
D^{K}_{R_{2}}(\varepsilon_{dep})=1-\log\left[(1-p)^{2}+1\right],
\end{equation}
\begin{equation}\label{Eq:DDEPT}
D^{K}_{T_{2}}(\varepsilon_{dep})=1-(1-p)^{2}.
\end{equation}
The value of the decohering power of the depolarizing channel with respect to the $l_{1}$-norm, the R$\acute{e}$nyi $2$-relative entropy, and the Tsallis relative $2$-relative only depend on the parameter $p$,i.e., the decohering power of the depolarizing channel is same to all reference basis. The variation of the decohering power of the depolarizing channel is depicted in Fig.\ref{Fig_3}

According to Eq.~\eqref{Eq:1K}, Eq.~\eqref{Eq:R2K}, Eq.~\eqref{Eq:T2K}, and Eq.~\eqref{Eq:CP2}, we have the cohering power of the depolarizing channel with respect ro the $l_{1}$-norm, the R$\acute{e}$nyi $2$-relative entropy, and the Tsallis $2$-relative entropy
\begin{equation}\label{Eq:CDEP}
C^{K}_{1}(\varepsilon_{dep})=C^{K}_{R_{2}}(\varepsilon_{dep})=C^{K}_{T_{2}}(\varepsilon_{dep})=0.
\end{equation}
It is obvious that the depolarizing channel has no cohering power with respect to arbitrary reference basis.

\subsection{Flip channels}

Finally, we study the dynamics of the flip channels $\varepsilon^{j}_{f}$, which can be described by~\cite{Nielsen00}
\begin{equation}\label{Eq:FD}
\varepsilon^{j}_{f}(\rho)=(1-p)\rho+p\sigma_{j}\rho\sigma_{j},
\end{equation}
where $j=x, y, z$ denotes the bit flip channel, bit-phase flip channel and the phase flip channel, respectively.

According to Eq.~\eqref{Eq:1K}, Eq.~\eqref{Eq:R2K}, Eq.~\eqref{Eq:T2K} and Eq.~\eqref{Eq:DP2}, the decohering power of the flip channels for the $l_{1}$-norm, the R$\acute{e}$nyi $2$-relative entropy and the Tsallis $2$-relative entropy are
\begin{equation}\label{Eq:DFC1}
D^{K}_{1}(\varepsilon^{j}_{f})=1-\min\limits_{\rho\in M}
\left\{\sqrt{4pr^{2}_{j}(1-p(1-k^{2}_{j}))+(1-2p)^{2}}\right\},
\end{equation}
\begin{equation}\label{Eq:DFCR}
D^{K}_{R_{2}}(\varepsilon^{j}_{f})=1-\min\limits_{\rho\in M}\left\{2\log\left[\frac{1}{2}\left(\sqrt{\tau}+\sqrt{\zeta}\right)\right]\right\},
\end{equation}
\begin{equation}\label{Eq:DFCT}
D^{K}_{T_{2}}(\varepsilon^{j}_{f})=2-\min\limits_{\rho\in M}
\left\{\left[\frac{1}{2}\left(\sqrt{\tau}+\sqrt{\zeta}\right)\right]^2\right\},
\end{equation}
where $\tau=1+(1-2p)^2+4pr_{j}(k_{j}-(1-p)r_{j})$, and $\zeta=1+(1-2p)^2-4pr_{j}(k_{j}+(1-p)r_{j}$. For the $x, y, z$ basis, $k_{z}$ is either 0 or 1. Then we have, for any $j\in\{x, y, z\}$
\begin{equation}\label{Eq:DFC11}
D^{K}_{1,x}(\varepsilon^{j}_{f})=D^{K}_{1,y}(\varepsilon^{j}_{f})=D^{K}_{1,z}(\varepsilon^{j}_{f})=1-|1-2p|,
\end{equation}
\begin{equation}\label{Eq:DFCR1}
D^{K}_{R_{2},x}(\varepsilon^{j}_{f})=D^{K}_{R_{2},y}(\varepsilon^{j}_{f})=D^{K}_{R_{2},z}(\varepsilon^{j}_{f})=1-\log(1+(1-2p)^{2}),
\end{equation}
\begin{equation}\label{Eq:DFCT1}
D^{K}_{T_{2},x}(\varepsilon^{j}_{f})=D^{K}_{T_{2},y}(\varepsilon^{j}_{f})=D^{K}_{T_{2},z}(\varepsilon^{j}_{f})=1-\log(1-2p)^{2}.
\end{equation}
\begin{figure}\label{fig4}
 \includegraphics[scale=0.6]{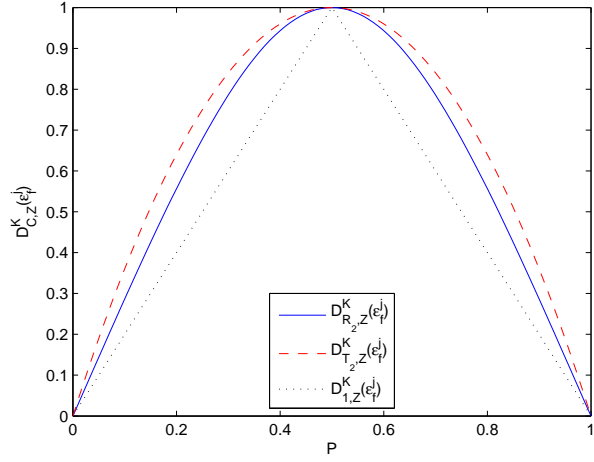}
\caption{(Color online)The decohering power of the flip channel on the $z$ basis under the $l_{1}$-norm(black points line), the R$\acute{e}$nyi $2$-relative entropy(blue valid line), and the Tsallis $2$-relative entropy(red dotted line denotes).}.
\label{Fig_4}
\end{figure}
\newline The decohering power of flip channels with respect to the $x, y, z$ basis can be seen in Fig. \ref{Fig_4}, from which we know that they all increased firstly and then decreased, reaching the maximum 1 at $p=\frac{1}{2}$.

Considering the cohering power of the flip channels, we taking Eq.\eqref{Eq:FD} into Eq.~\eqref{Eq:1K}, Eq.~\eqref{Eq:R2K}, Eq.~\eqref{Eq:T2K} respectively, then using Eq.~\eqref{Eq:CP2}, we have
\begin{equation}\label{CFC1}
C^{K}_{1}(\varepsilon^{j}_{f})=2pk_{j}\sqrt{1-k^{2}_{j}},
\end{equation}
\begin{equation}\label{CFCR}
C^{K}_{R_{2}}(\varepsilon^{j}_{f})=2\log\left(\sqrt{(1-p)^{2}+p(2-p)k^{2}_{j}}+p\sqrt{1-k^{2}_{j}}\right),
\end{equation}
\begin{equation}\label{CFCT}
C^{K}_{T_{2}}(\varepsilon^{j}_{f})=\left[\sqrt{(1-p)^{2}+p(2-p)k^{2}_{j}}+p\sqrt{1-k^{2}_{j}}\right]^{2}-1.
\end{equation}
The maximal cohering power of the flip channels with respect to the $l_{1}$-norm, the R$\acute{e}$nyi $2$-relative entropy and the Tsallis $2$-relative entropy is 1, respectively. For the $x, y, z$ basis, $k_{j}$ is 0 or 1, then we have, for any $j\in\{x, y, z\}$
\begin{equation}\label{CFC1xyz}
C^{K}_{1,x}(\varepsilon^{j}_{f})=C^{K}_{1,y}(\varepsilon^{j}_{f})=C^{K}_{1,z}(\varepsilon^{j}_{f})=0,
\end{equation}
\begin{equation}\label{CFCRxyz}
C^{K}_{R_{2},x}(\varepsilon^{j}_{f})=C^{K}_{R_{2},y}(\varepsilon^{j}_{f})=C^{K}_{R_{2},z}(\varepsilon^{j}_{f})=0,
\end{equation}
\begin{equation}\label{CFCTxyz}
C^{K}_{T_{2},x}(\varepsilon^{j}_{f})=C^{K}_{T_{2},y}(\varepsilon^{j}_{f})=C^{K}_{T_{2},z}(\varepsilon^{j}_{f})=0.
\end{equation}
That is to say the flip channels doesn't have any cohering power with respect to the $x, y, z$ basis.

\section{conclusion}\label{sec:conclusion}

Quantum coherence plays an important role in quantum information procession. In this paper, We mainly introduce the cohering and decohering power of the Markovian channels in terms of the $l_{1}$-norm, the R$\acute{e}$nyi $\alpha$-relative entropy and the Tsallis $\alpha$-relative entropy. For convenience, we calculate the special case of $\alpha=2$. We find that the decohering power for the amplitude channel, the phase damping channel, and the depolarizing channel with respect to the three measures we investigated are increased monotonically with $p$ from 0 to 1. But for the flip channels it increases when $p\in[0, \frac{1}{2}]$ and decreases when $p\in[\frac{1}{2}, 1]$, reaching a maximal value of 1 at $p=\frac{1}{2}$. Moreover, the maximal cohering power of the amplitude damping channel and the flip channels are the same to 1. However, the depolarizing channel has no cohering power with respect to arbitrary basis. These results may be useful to the study of coherence.

While we only study the cohering and decohering power of the Markovian channels for one-qubit states with respect to the $l_{1}$-norm, the R$\acute{e}$nyi $\alpha$-relative entropy and the Tsallis $\alpha$-relative entropy in the case of $\alpha=2$ . The cohering and decohering power of one-qubit channels for the general parameters $\alpha$ need further investigated.

\end{document}